\documentclass[12pt]{article}
\usepackage[dvips]{graphicx}
\def\doublespace{\lineskip      .25 ex\baselineskip 3.0
ex\lineskiplimit 0 ex\parskip 1.0 ex plus.50 ex minus .25 ex}%
\def\bea{\begin{eqnarray}}
\def\eea{\end{eqnarray}}
\def\be{\begin{equation}}
\def\ee{\end{equation}}

\def\m\mu
\def\n{\nu}

\begin{document}
\doublespace

\title{Searching Exact Solutions for Compact\\ Stars in Braneworld: a conjecture}
\author{ J.
Ovalle\footnote{jovalle@usb.ve; jovalle@fisica.ciens.ucv.ve}
 \\
\vspace*{.25cm}\\
 Escuela de F\'{\i}sica, Universidad Central de Venezuela, \\ Caracas,
 Venezuela.
 }
\date{}
\maketitle
\begin{abstract}

In the context of the braneworld, a method to find consistent
solutions to Einstein's field equations in the interior of a
spherically symmetric, static and non uniform stellar distribution
with Weyl stresses is developed. This method, based in the fact
that any braneworld stellar solution must have the general
relativity solution as a limit, produces a constraint which
reduces the degrees of freedom on the brane. Hence the non
locality and non closure of the braneworld equations can be
overcome. The constraint found is physically interpreted as a
necessary condition to regain general relativity, and a particular
solution for it is used to find an exact and physically acceptable
analytical internal solution to no-uniform stellar distributions
on the brane. It is shown that such an exact solution is possible
due to the fact that bulk corrections to pressure, density and a
metric component are a null source of anisotropic effects on the
brane. A conjecture is proposed about the possibility of finding
physically relevant exact solutions to non-uniform stellar
distributions on the brane.

keywords: Braneworld; Gravity in extra dimensions; General
relativity.

\end{abstract}
\newpage
\section{Introduction}

The standard model and the general relativity represents the two
great theories in fundamental physics. The success of general
relativity is beyond any doubt, however due to its inconsistency
with quantum mechanics, it is not possible to ensure that this
theory keeps its original structure at high energies. Indeed it is
expected that such a structure will be modified and thus general
relativity will emerge as the low energy limit of a more general
theory not yet discovered. In spite of an enormous effort, it has
not been possible to construct an unified theory which describes
all fundamental interactions. To carry out this project, a
consistent quantum theory to gravity must be cons\-tructed.
Fortunately there is a theory which seems to lead to the expected
quantum gravity.

The superstring/M-theory is considered one of the most promising
candidate theories of quantum gravity. It describes gravity as a
high dimensional interaction which becomes effectively four
dimensional at low enough ener\-gies. This theory has inspired the
construction of braneworld models, in which the standard model
gauge fields are confined to our observable universe (the brane),
while gravity propagates in all spatial dimensions (the bulk)
\cite{rs}.

The implications of the braneworld theory on general relativity
have been extensively investigated \cite{chr}-\cite{cosmox} (see
also a review paper \cite{maartRev2004} and references therein),
most of theme on cosmological scenarios
\cite{bdel00}-\cite{cbkt07} (see also \cite{cosmo04} and
references therein). The studies on astrophysics consequences
\cite{germ}-\cite{baff} are mostly limited to exterior solutions,
even though it is well known that gravitational collapse could
produce very high energies, where the braneworld corrections to
general relativity would become significant
\cite{twism}-\cite{gergely2007}. Unfortunately there has not been
reported any consistent internal stellar solution \footnote{In the
pioneer work of Germani and Maartens \cite{germ} exact solutions
for a uniform distribution were found, where it can be seen how
much more difficult it would be to find a solution except for
uniform stellar distributions.}, therefore the braneworld
consequences on physical processes in the interior remain unknown
so far.

As is well known, it is extremely complicated to find exact
interior solutions in general relativity \cite{kramer}. Even for
the simple case of a static perfect fluid, only very few reported
exact solutions are of physical interest \cite{DK98}. Hence it is
not surprising that finding exact solutions in the context of
braneworld, where new terms from the bulk have to be considered,
becomes a very complicated process, except when uniform
distributions are considered. \footnote{Recently some braneworld
wormhole solutions were reported \cite{flobo07}, where an exact
dust solution can be clearly seen.} Consequently important issues,
as the role played by the density gradients as a source of Weyl
stresses, and its importance in different physical processes in
the interior, have not been elucidated yet. An exact solution
would be useful to carry out an analytic analysis. Unfortunately
there has not been possible to construct any exact internal
solution to non-uniform stellar distributions in the context of
the braneworld. The reason is that the non locality and non
closure of the braneworld equations, produced by the projection of
the bulk Weyl tensor on the brane, lead to a very complicated
system of equations which difficult the study of non-uniform
distributions. Indeed, there is not any well established criteria
about what restriction should be considered on braneworld
equations to obtain a closed system \footnote{In \cite{shtanov07}
a closed system of equations for static spherically symmetric
situation in the case of ${\cal P}=0$ was studied.}. To solve this
issue, it is necessary a better understanding of the bulk geometry
and how our 4D spacetime is embedded.

As already mentioned, the study on density gradient effects in the
interior remain unknown so far. For instance, the role played by
density gradients during the gravitational collapse represents an
open issue which has not been studied in the context of
braneworld. In this respect, it is well known that the collapse of
a homogeneous star leads to a a non-static exterior solution
\cite{bgm}-\cite{kp2004}. Moreover, Govender and Dadhich
\cite{mgnd} proved that a collapsing sphere on the brane radiates,
and they shown that the exterior for this radiative sphere can be
described by a Vaidya metric that envelops the collapsing region.
This represents a new and complete different aspect from the
four-dimensional general relativity situation, and a remarkable
issue which should be deeper investigated. In this context, it
would be interesting to elucidate the effects produced by density
gradients as a source of Weyl stresses, which eventua\-lly will
be, together with the non-vanishing of the effective pressure at
the stellar surface, the responsible for the non-static nature of
the exterior solution.

In this paper we develop a method which helps to overcome the non
locality and non closure of the braneworld equations when a
non-uniform stellar distribution is considered. This approach is
based in the fact that any braneworld stellar solution must have
the general relativity solution as a limit. This statement is
exploited in this paper to generate a constraint which helps to
reduce the degrees of freedom for an non-uniform distribution
having local and non-local bulk terms on the brane. A particular
solution for this constraint is used to find an exact and
physically acceptable internal solution to no-uniform stellar
distributions on the brane.

This paper is organized as follows. In Section 2 the Einstein's
field equations and matching conditions in the brane for a
spherically symmetric distribution is reminded. In Section 3 a toy
solution is used to show how the general relativity limit could be
easily lost when a solution is achieved on the brane. In Section 4
the source of the general relativity limit problem is clearly
identified and a solution for this is given. This solution shows
an explicit constraint emerging as a natural consequence of the
general relativity solution. In Section 5 the constraint
identified in the previous section is used to find an exact
analytical solution to a no-uniform distribution with bulk terms
on the brane. In Section 6 an exact and physically acceptable
solution is found. It is shown that this exact solution can be
obtained due to the null consequence of bulk corrections to
pressure, density and a metric component ($p$, $\rho$ and $\nu$)
to produce anisotropic effects on the brane. In the last section
the conclusions are presented and a conjecture is proposed about
the possibility of finding physically relevant exact solutions to
non-uniform stellar distributions on the brane.

\section{The field equations and matching conditions}

The Einstein field equations on the brane may be written as a
modification of the standard field equations \cite{germ2}
\begin{equation}\label{einst}
G_{\mu\nu}=-8\pi T_{\mu\nu}^{T}-\Lambda g_{\mu\nu},
\end{equation}
where $\Lambda$ is the cosmological constant on the brane. The
energy-momentum tensor has new terms carrying bulk effects onto
the brane:
\begin{equation}\label{tot}
T_{\mu\nu}\rightarrow T_{\mu\nu}^{\;\;T}
=T_{\mu\nu}+\frac{6}{\sigma}S_{\mu\nu}+\frac{1}{8\pi}{\cal
E}_{\mu\nu},
\end{equation}
here $\sigma$ is the brane tension. The new terms $S_{\mu\nu}$ and
$\cal{E}_{\mu\nu}$ are the high-energy corrections and $KK$
corrections respectively, and are given by
\begin{equation}\label{s}
S_{\mu\nu}=\frac{1}{12}T_\alpha^{\;\alpha}T_{\mu\nu}-\frac{1}{4}T_{\mu\alpha}T^\alpha_{\;\nu}
+\frac{1}{24}g_{\mu\nu}\left[3T_{\alpha\beta}T^{\alpha\beta}-(T_\alpha^{\;\alpha})^2\right],
\end{equation}
\begin{equation}\label{e}
-8\pi{\cal E}_{\mu\nu}=-\frac{6}{\sigma}\left[{\cal U}(u_\mu
u_\nu+\frac{1}{3}h_{\mu\nu})+{\cal P}_{\mu\nu}+{\cal
Q}_{(\mu}u_{\nu)}\right],
\end{equation}
being ${\cal U}$ the dark radiation and ${\cal P}_{\mu\nu}$ and
${\cal Q}_\mu $ the anisotropic stress and energy flux
respectively.

We consider a spherically symmetric static distribution, hence
$Q_\mu =0$ and
\begin{equation}
{\cal P}_{\mu\nu}={\cal P}(r_\mu r_\nu+\frac{1}{3}h_{\mu\nu}),
\end{equation}
where $r_\mu$ is a unit radial vector and
$h_{\mu\nu}=g_{\mu\nu}-u_\mu u_\nu$ the projection tensor with
4-velocity $u^{\mu}$. The line element is given in
Schwarzschild-like coordinates by
\begin{equation}
\label{metric}ds^2=e^\nu dt^2-e^\lambda dr^2-r^2\left( d\theta
^2+\sin {}^2\theta d\phi ^2\right)
\end{equation}
where $\nu$ and $\lambda$ are functions of $r$.

The metric (\ref{metric}) has to satisfy (\ref{einst}). In our
case with $\Lambda=0$ we have:

\begin{equation}
\label{ec1}-8\pi \left( \rho
+\frac{1}{\sigma}\left(\frac{\rho^2}{2}+\frac{6}{k^4}\cal{U}\right)
\right) =-\frac 1{r^2}+e^{-\lambda }\left( \frac
1{r^2}-\frac{\lambda _1}r\right),
\end{equation}

\begin{equation}
\label{ec2}-8\pi
\left(-p-\frac{1}{\sigma}\left(\frac{\rho^2}{2}+\rho p
+\frac{2}{k^4}\cal{U}\right)-\frac{4}{k^4}\frac{\cal{P}}{\sigma}\right)
=-\frac 1{r^2}+e^{-\lambda }\left( \frac 1{r^2}+\frac{\nu
_1}r\right),
\end{equation}

\begin{eqnarray}
\label{ec3}-8\pi
\left(-p-\frac{1}{\sigma}\left(\frac{\rho^2}{2}+\rho p
+\frac{2}{k^4}\cal{U}\right)+\frac{2}{k^4}\frac{\cal{P}}{\sigma}\right)
= \frac 14e^{-\lambda }\left[ 2\nu _{11}+\nu _1^2-\lambda _1\nu
_1+2 \frac{\left( \nu _1-\lambda _1\right) }r\right] \nonumber \\,
\end{eqnarray}

\begin{equation}
\label{con1}p_{1}=-\frac{\nu_1}{2}(\rho+p),
\end{equation}
where $f_1\equiv df/dr$ and $k^2=8{\pi}$. The general relativity
is regained when $\sigma^{-1}\rightarrow 0$ and (\ref{con1})
becomes a lineal combination of (\ref{ec1})-(\ref{ec3}).


The Israel-Darmois matching conditions at the stellar surface
$\Sigma$ give
\begin{equation}
\label{matching1} [G_{\mu\nu}r^\nu]_{\Sigma}=0
\end{equation}
where $[f]_{\Sigma}\equiv f(r)\mid_{R^+}-f(r)\mid_{R^-}$ Using
(\ref{matching1}) and the field equation (\ref{einst}) with
$\Lambda=0$ we have
\begin{equation}
\label{matching2} [T^{\;\;T}_{\mu\nu}r^\nu]_{\Sigma}=0,
\end{equation}
which leads to
\begin{equation}
\label{matching3} \left[
\left(p+\frac{1}{\sigma}\left(\frac{\rho^2}{2}+\rho p
+\frac{2}{k^4}\cal{U}\right)+\frac{4}{k^4}\frac{\cal{P}}{\sigma}\right)
\right]_{\Sigma}=0.
\end{equation}
This takes the final form
\begin{equation}
\label{matchingf}
p_R+\frac{1}{\sigma}\left(\frac{\rho_R^2}{2}+\rho_R p_R
+\frac{2}{k^4}{\cal U}_R^-\right)+\frac{4}{k^4}\frac{{\cal
P}_R^-}{\sigma} = \frac{2}{k^4}\frac{{\cal
U}_R^+}{\sigma}+\frac{4}{k^4}\frac{{\cal P}_R^+}{\sigma},
\end{equation}
where $f_R\equiv f(r)\mid_{r=R}$. The equation (\ref{matchingf})
gives the general matching condition for any static spherical star
on braneworld \cite{germ} \footnote{The general matching
conditions on the brane for a spherically symmetric vacuum region
embedded into a cosmological environment can be seen in
\cite{gergely2005}-\cite{gergely2007-2}}. When
$\sigma^{-1}\rightarrow 0$ we obtain the well known matching
condition $p_R =0$. In the particular case of the Schwarzschild
exterior solution ${\cal U}^+={\cal P}^+ =0$, the matching
condition (\ref{matchingf}) becomes:
\begin{equation}
\label{matchingfS}
p_R+\frac{1}{\sigma}\left(\frac{\rho_R^2}{2}+\rho_R p_R
+\frac{2}{k^4}{\cal U}_R^-\right)+\frac{4}{k^4}\frac{{\cal
P}_R^-}{\sigma} = 0.
\end{equation}
It is easily seen that the matching conditions do not have a
unique solution on the brane.

\section{Losing general relativity}

In the case of a non-uniform static distribution with local and
non-local bulk terms, we have an indefinite system of equations
(\ref{ec1})-(\ref{con1}) whose solution must be determined with
the add of additional information. However it is not clear what
kind of restriction should be considered. As mentined earlier,
this is an open problem which solution imply more information of
the five dimensional geometry and a better understanding of how
the brane is embedded in the bulk. Hence the study of any method
which leads to a reduction of the degrees of freedom on the brane
represents a subject of great interest. In order to construct a
method, let us start from the simplest way to find a solution to
the system (\ref{ec1})-(\ref{con1})

From the field equations (\ref{ec2}) and (\ref{ec3}) we obtain
\begin{equation}
\label{pp}\frac{8\pi}{k^4}\frac{{\cal
P}}{\sigma}=\frac{1}{6}\left(G_1^1-G_2^2\right)
\end{equation}
and
\begin{equation}
\label{uu}\frac{6}{k^4}\frac{{\cal
U}}{\sigma}=-\frac{3}{\sigma}\left(\frac{\rho^2}{2}+\rho\,
p\right)+\frac{1}{8\pi}\left(2G_2^2+G_1^1\right)-3p,
\end{equation}
with
\begin{equation}
\label{g11} G_1^1=-\frac 1{r^2}+e^{-\lambda }\left( \frac
1{r^2}+\frac{\nu _1}r\right),
\end{equation}
\begin{equation}
\label{g22} G_2^2=\frac 14e^{-\lambda }\left[ 2\nu _{11}+\nu
_1^2-\lambda _1\nu _1+2 \frac{\left( \nu _1-\lambda _1\right)
}r\right].
\end{equation}
Now using (\ref{uu}) in the field equation (\ref{ec1}) we have
\begin{eqnarray}
\label{e1g}
-\lambda_1e^{-\lambda}+e^{-\lambda}\left(\frac{\nu_{11}+\nu_1^2/2+2\nu_1/r+2/r^2}{\nu_1/2+2/r}\right)=\frac{2}{r^2(\nu_1/2+2/r)}
\nonumber \\
\nonumber \\
-8\pi\frac{\left(\rho-3p-\frac{1}{\sigma}\rho
(\rho+3p)\right)}{(\nu_1/2+2/r)},
\end{eqnarray}
which formal solution is
\begin{eqnarray}
\label{primsol}
e^{-\lambda}=e^{-I}\left(\int_0^r\frac{e^I}{(\frac{\nu_1}{2}+\frac{2}{r})}\left[\frac{2}{r^2}-8\pi(\rho-3p-\frac{1}{\sigma}\left(\rho^2
+3\rho\, p)\right)\right]dr+c\right),
\end{eqnarray}
\begin{eqnarray}
\label{I} I\equiv
\int\frac{(\nu_{11}+\frac{\nu_1^2}{2}+\frac{2\nu_1}{r}+\frac{2}{r^2})}{(\frac{\nu_1}{2}+\frac{2}{r})}dr.
\end{eqnarray}
Hence finding $\nu$, $\rho$ and $p$ satisfying (\ref{con1}) we
would be able to find $\lambda$, ${\cal P}$ and ${\cal U}$ by
(\ref{primsol}), (\ref{pp}) and (\ref{uu}) respectively. Therefore
finding a solution to the system (\ref{ec1})-(\ref{con1}), at
least from the mathematical point of view, seems not very
complicated. We will see that even from a pure mathematical point
of view, to find a consistent solution is not easy at all. To
address this we consider a simple solution to (\ref{con1}) showed
below
\begin{equation}
\label{solii}
\rho=A+Br^{k/2};\;\;\;p=-A-\frac{B}{2}r^{k/2};\;\;\;e^{\nu/2}=Cr^{k/2},
\end{equation}
where $k>0$, $A$, $B$ and $C$ are constants. Although such a
solution is not of physical interest, it will be useful in showing
the problem which arises when the complete solution is found by
this method. In this sense, (\ref{solii}) is considered to be just
a toy solution.

Using (\ref{solii}) in (\ref{primsol}) we should expect to find a
solution given by
\begin{eqnarray}
\label{expect}  e^{-\lambda}=1-\frac{8\pi}{r}\int_0^r r^2\rho
dr+\frac{1}{\sigma}(Bulk\;\;effects),
\end{eqnarray}
and then taking the $\frac{1}{\sigma}\rightarrow 0$ limit the
general relativity would be regained. Unfortunately this does not
happen, thus the solution found by the simple way showed here is
not even mathematically consistent, that is the solution evaluated
at $\frac{1}{\sigma}=0$ is not a solution to
(\ref{ec1})-(\ref{ec3}) at $\frac{1}{\sigma}=0$. Hence the
apparently simplest and straightforward way lead us to an
inconsistent solution, so obviously there is something incorrect
in the way we just used to obtain $\lambda(r)$, therefore it is
necessary to carry out a more careful analysis. This will be
addressed in the next section.

\section{Recovering the general relativity limit}

In order to find the source of the "general relativity limit
problem" found in the previous section, let us write the field
equation (\ref{ec1}) as usual in general relativity
\begin{equation}
\label{usual} e^{-\lambda}=1-\frac{8\pi}{r}\int_0^r
r^2\tilde{\rho}dr,
\end{equation}
where
\begin{equation}
\label{deneffec} \tilde{\rho}\equiv\rho
+\frac{1}{\sigma}\left(\frac{\rho^2}{2}+\frac{6}{k^4}\cal{U}\right).
\end{equation}
Thus we can see the high-energy effects of bulk gravity and
nonlocal corrections from the bulk Weyl curvature. We can see that
the "solution" (\ref{usual}) depends itself on $\lambda(r)$ and
$\lambda_1(r)$ through $\cal{U}$ ({\ref{uu})-({\ref{g22}), hence
it it represents an integral differential equation for
$\lambda(r)$. Now we can realize why the formal solution
(\ref{primsol}) cannot written by the way showed in
(\ref{expect}). Such solution has mixed up general relativity
terms with bulk effects on the brane. In consequence finding how
the general relativity limit problem originates becomes very
complicated. To clarify this point let us rewrite the differential
equation (\ref{e1g}) by
\begin{eqnarray} \label{edlrw}
\left[\frac{-\lambda_1e^{-\lambda}}{r}+\frac{e^{-\lambda}}{r^2}-\frac{1}{r^2}+8\pi
\rho\right]+\left[-\lambda_1e^{-\lambda}(\frac{\nu_1}{2}+\frac{1}{r})+\right.
\nonumber \\
\nonumber \\
\left.e^{-\lambda}(\nu_{11}+\frac{\nu_1^2}{2}+2\frac{\nu_1}{r}+\frac{1}{r^2})-\frac{1}{r^2}-8\pi
3p-\frac{8\pi}{\sigma}\rho\left(\rho+3p\right)\right]=0,
\end{eqnarray}
here the left bracket has the standard general relativity terms
and the right one the high energy terms with bulk Weyl curvature
contribution to the differential equation of $\lambda(r)$. It is
worth noticing that in the right bracket only high energy terms
are manifestly bulk contributions, hence to keep bulk Weyl
curvature contributions under control when the formal solution
(\ref{primsol}) is achieved is not at all easy.

We propose a solution to (\ref{edlrw}) which can be written in a
way that allows to see clearly the effects of bulk contribution on
the brane
\begin{equation}
\label{edlrwss} e^{-\lambda}={1-\frac{8\pi}{r}\int_0^r r^2\rho
dr}+e^{-I}\int_0^r\frac{e^I}{(\frac{\nu_1}{2}+\frac{2}{r})}\left[H(p,\rho,\nu)+\frac{8\pi
}{\sigma}\left(\rho^2+3\rho p\right)\right]dr,
\end{equation}
where $I$ is given once again by (\ref{I}) and
\begin{equation}
\label{finalsol} H(p,\rho,\nu)\equiv 8\pi
3P-\left[\mu_1(\frac{\nu_1}{2}+\frac{1}{r})+\mu(\nu_{11}+\frac{\nu_1^2}{2}+\frac{2\nu_1}{r}+\frac{1}{r^2})-\frac{1}{r^2}\right],
\end{equation}
with
\begin{equation}
\label{def} \mu\equiv 1-\frac{8\pi}{r}\int_0^r r^2\rho dr.
\end{equation}
The reader may check that (\ref{edlrwss}) satisfies (\ref{edlrw}).

It is easy to see by (\ref{ec1})-(\ref{ec3}) that at
$\sigma^{-1}=0$ we have
\begin{equation}
\label{cond} 2G_2^2+G_1^1=\left[\mu_1(\frac{\nu_1}{2}+\frac{1}{r})
+\mu(\nu_{11}+\frac{\nu_1^2}{2}+\frac{2\nu_1}{r}+\frac{1}{r^2})-\frac{1}{r^2}\right]=8\pi
3P,
\end{equation}
so the function $H(p,\rho,\nu)$ vanishes when
$\sigma^{-1}\rightarrow 0$ and thus it may be interpreted as a
function which measures the anisotropic effects due to bulk
consequences to $p$, $\rho$ and $\nu$. Since $H(p,\rho,\nu)$
vanishes when $\sigma^{-1}\rightarrow 0$ the solution
(\ref{edlrwss}) seems having the well known general relativity
limit:
\begin{eqnarray}
\label{s0gr}
e^{-\lambda}|_{\sigma^{-1}=0}=1-\frac{8\pi}{r}\int_0^r r^2\rho dr.
\end{eqnarray}
However using our toy solution (\ref{solii}) we can see that
$H(p,\rho,\nu)$ does not vanish. We should expect this to be due
to the fact that this solution was built using only (\ref{con1}),
without any additional assumption. Hence we may learn that there
is something else to be considered to find at least a
mathematically consistent solution. In order to close the system
(\ref{ec1})-(\ref{con1}), at least two conditions must be imposed.
Thus we could use $H(p,\rho,\nu)=0$ as a convenient constraint
which lets us make sure that $\lambda(r)$ has the general
relativity solution as a limit. In any case every solution found
must satisfy the condition $H(p,\rho,\nu)\rightarrow 0$ when
$\sigma^{-1}\rightarrow 0$. In that sense this limit could be
thought of as a constraint itself which might be useful to close
the system. However we cannot forget that the $H(p,\rho,\nu)$
contribution to $\lambda(r)$ is under integration, therefore the
constraint to be considered is
\begin{equation}
\label{constraint} lim_{\sigma^{-1}\rightarrow\;\;
0}\;\;\int_0^r\frac{e^I}{(\frac{\nu_1}{2}+\frac{2}{r})}H(p,\rho,\nu)dr
=0,
\end{equation}
and it must be satisfied as a necessary condition to regain
general relativity. Since the constraint (\ref{constraint}) is
proportional to $H(p,\rho,\nu)$, we may interpret this constraint
as the continuous elimination of deformations produced on general
relativity by anisotropic effects generated for bulk corrections
to $p$, $\rho$ and $\nu$.

The method shown here leaves us with three unknowns $\rho$, $p$
and $\nu$ sa\-tisfying (\ref{con1}), and eventually the constraint
(\ref{constraint}). However there is something which we might be
concerned about. We can see that Eq. (\ref{con1}) does not have
any manifestly bulk contribution. Thus it might be thought that
the solution eventually found for $p$, $\rho$ and $\nu$ would be
the same as the general relativity one. Hence there would not be
any bulk contribution for these functions. Actually the fact that
Eq. (\ref{con1}) does not have any manifestly bulk terms does not
mean its solution does not either. Indeed the bulk contribution
for $p$, $\rho$ and $\nu$ will appear by means of matching
conditions, where the assumption of vanishing pressure will be
dropped \cite{deru}, \cite{gergely2007} as will be shown in the
next section.

\section{A solution}

We shall construct a simple mathematically consistent solution in
order to clarify the method described in the previous section.
First of all let us enforce the constraint
\begin{equation}
\label{constraint2} H(p,\rho,\nu)=0
\end{equation}
on the brane to make sure we have a solution for the geometric
function $\lambda(r)$ with the correct limit. Of course
(\ref{constraint2}) satisfies (\ref{constraint}) and what it means
is that eventual bulk corrections to $p$, $\rho$ and $\nu$ will
not produce anisotropic effects on the brane. Secondly, to
evaluate (\ref{I}) and eventually (\ref{edlrwss}), we need to
consider a simple enough expression to $\nu(r)$. Accordingly we
impose
\begin{equation}
\label{solmetr} e^{\nu}=Ar^{4}.
\end{equation}
A simple but unique solution to (\ref{con1}) and
(\ref{constraint2}) using (\ref{solmetr}) is given by
\begin{equation}\label{solsing}
\rho(r)=Br^{-4/3};\;\;\;\;\;\;8\pi p(r) =\frac{4}{r^2}-8\pi
3Br^{-4/3},
\end{equation}
where $A$ and $B$ are constants to be determined by matching
conditions.

Using (\ref{solmetr}) and (\ref{solsing}) in  (\ref{edlrwss}) we
obtain
\begin{equation}
e^{-\lambda}=1-\frac{2M}{r}\left(\frac{r}{R}\right)^{5/3}+\frac{8\pi
B}{\sigma}\left(\frac{18}{13}r^{-4/3}-\frac{12B}{7}r^{-2/3}\right),
\end{equation}}
where
\begin{equation}
\label{usual2} M=\int_0^R 4\pi r^2{\rho}dr,
\end{equation}
and where $R$ is the radius of the distribution. We see that the
solution thus generated is not regular at the origin, caused by
the imposed function (\ref{solmetr}) and the particular solution
(\ref{solsing}), but not due to the method described here. Finally
using (\ref{pp}) and (\ref{uu}) we obtain
\begin{equation}
\label{UandP} 4{\cal
P}=\frac{B}{r^{8/3}}\left(\frac{9}{26\pi}\frac{1}{r^{2/3}}-\frac{16}{17}B\right);
\;\;\;4{\cal
U}=\frac{B}{r^{8/3}}\left(\frac{1}{26\pi}\frac{1}{r^{2/3}}-\frac{3}{17}B\right).
\end{equation}

The bulk contribution to $p$, $\rho$ and $\nu$ can be found by
matching conditions. For instance let us consider the
Schwarzschild exterior solution
\begin{equation}
\label{SchwExt}
e^{\nu^{+}}=e^{-\lambda^{+}}=1-\frac{2\cal{M}}{r};\;\;\;\;\;\;\;{\cal
U}^{+}={\cal P}^{+}=0.
\end{equation}
Considering the matching condition $[ds^2]_{\Sigma}=0$ at the
stellar surface $\Sigma$ we have
\begin{equation}
\label{matchSchw1} AR^{4}=1-\frac{2\cal{M}}{R},
\end{equation}
\begin{equation}
\label{matchSchw2} \frac{2\cal{M}}{R}=\frac{2M}{R}-\frac{8\pi
B}{\sigma}\left(\frac{18}{13}R^{-4/3}-\frac{12B}{7}R^{-2/3}\right)
\end{equation}
and using (\ref{matchingfS}) we obtain
\begin{equation}
\label{matchSchw3}
B(\sigma)=\frac{\frac{1}{\sigma}765-2652{\pi}R^2+\sqrt{51}\sqrt{\frac{1}{{\sigma}^2}11475+\frac{1}{\sigma}28600{\pi}R^2+137904{\pi^2}R^4}}{\frac{1}{\sigma}6240{\pi}R^{2/3}},
\end{equation}
thus the bulk contribution to $p$, $\rho$ and $\nu$ can be seen
clearly through $A(\sigma)$ and $B(\sigma)$. Considering lineal
terms in $\sigma^{-1}$ we have
\begin{eqnarray}
\label{matchSchwf}
B(\sigma)=\frac{1}{6\pi{R^{2/3}}}+\frac{1}{\sigma}\frac{245}{15912\pi^2R^{8/3}}+{\cal
O}(\sigma^{-2}), \\
\nonumber \\
{\cal
M}(\sigma)=M-\frac{1}{\sigma}\frac{4}{R}\left(\frac{3}{13}-\frac{1}{21\pi}\right)+{\cal
O}(\sigma^{-2}), \\
\nonumber \\
A(\sigma)=\frac{1}{R^{4}}\left[1-\frac{2M}{R}+\frac{1}{\sigma}\frac{8}{R^2}\left(\frac{3}{13}-\frac{1}{21\pi}\right)\right]+{\cal
O}(\sigma^{-2}).
\end{eqnarray}
Thus we obtain our solution at first order in $1/\sigma$, where
the bulk effects on $p$, $\rho$ and $\nu$ can be easily seen by
\begin{equation}
\label{presford} p =\frac{4}{8\pi\,r^2}-
3\left[\frac{1}{6\pi{R^{2/3}}}+\frac{1}{\sigma}\frac{245}{15912\pi^2R^{8/3}}\right]r^{-4/3},
\end{equation}
\begin{equation}
\label{denford}
\rho=\left[\frac{1}{6\pi{R^{2/3}}}+\frac{1}{\sigma}\frac{245}{15912\pi^2R^{8/3}}\right]r^{-4/3},
\end{equation}
\begin{equation}
\label{nuford}
e^{\nu}=\left[1-\frac{2M}{R}+\frac{1}{\sigma}\frac{8}{R^2}\left(\frac{3}{13}-\frac{1}{21\pi}\right)\right]\left(\frac{r}{R}\right)^{4},
\end{equation}
leaving
\begin{equation}
\label{Pford} \frac{4{\cal P}}{\sigma}
=\frac{1}{{\pi^2}R^{2/3}r^{8/3}}\left[\frac{3}{52r^{2/3}}-\frac{4}{153R^{2/3}}\right]\frac{1}{\sigma},
\end{equation}
\begin{equation}
\label{Uford} \frac{{\cal U}}{\sigma}
=\frac{1}{48{\pi^2}R^{2/3}r^{8/3}}\left[\frac{1}{13r^{2/3}}-\frac{1}{17R^{2/3}}\right]\frac{1}{\sigma},
\end{equation}
and
\begin{equation}
\label{fin}
e^{-\lambda}=1-\frac{2M}{r}\left(\frac{r}{R}\right)^{5/3}+\frac{1}{\sigma}\frac{8}{R^{2/3}r^{2/3}}\left(\frac{3}{13}\frac{1}{r^{2/3}}-
\frac{1}{21{\pi}R^{2/3}}\right).
\end{equation}

The solution (\ref{presford})-(\ref{fin}) represents an exact
analytical solution to the system (\ref{ec1})-(\ref{con1}) with
the correct limit at low energies. The algorithm developed to
obtain this solution runs as follows:

\begin{itemize}

\item Step 1: Impose the constraint $H(p,\rho,\nu)=0$ to make sure
we have a solution for the geometric function $\lambda(r)$ with
the correct limit at low energies.

\item Step 2: Pick a simple enough expression to $\nu(r)$ to
obtain an analytical expression to the Eq. (\ref{I}).

\item Step 3: Use $\nu(r)$ in both the conservation equation
$\;p'=-\frac{\nu'}{2}(\rho+p)\;$ and the constraint
$H(p,\rho,\nu)=0$ to find $p(r)$ and $\rho(r)$.

\item Step 4: Use $p(r)$, $\rho(r)$ and $\nu(r)$ in
(\ref{edlrwss}) to obtain $\lambda(r)$.

\item Step 5: Find ${\cal P}$ and ${\cal U}$ using (\ref{pp}) and
(\ref{uu}).

\item Step 6: Drop out the condition of vanishing pressure at the
surface to obtain the bulk effect on any constant $C\rightarrow
C(\sigma)$. Then we are able to find the effects of bulk gravity
on pressure and density.

\end{itemize}

It is worth noticing that this algorithm is general and works even
for a numerical analysis, where any arbitrary expression for
$\nu(r)$ at step 2 can be chosen.

So far the solution found is a simple mathematically consistent
solution. It was constructed to clarify the method described in
the previous section and summarized in the algorithm shown here.
In order to obtain a physically acceptable solution, it is
necessary to carry out a more careful analysis. This will be
addressed in the next section.

\section{A physically acceptable solution}

The approach developed here ensures general relativity as a limit,
but it does not say anything about finding a physically relevant
solution. Of course it is not enough to pick out a simple
expression to $\nu(r)$. Indeed choosing a convenient expression
for $\nu(r)$, leading to a physically acceptable solution,
represents a critical issue. In this paper it was found that the
following expression for the metric component
\begin{equation}
\label{regularmet00} e^{\nu}=A(1+Cr^{2})^3
\end{equation}
leads to a simple and physically acceptable solution for
(\ref{con1}) and (\ref{constraint2}) given by
\begin{equation}\label{regularpress}
8\pi\,p(r)=\frac{9\,
    C\, \left( 1 - C\,
        r^2 \right) }{2\, {\left( 1 + C\, r^2 \right) }^2},
\end{equation}
and
\begin{equation}\label{regularden}
8\pi\,\rho(r)=\frac{3\,
    C\, \left( 3 + C\,
        r^2 \right) }{2 {\left( 1 + C\, r^2 \right)
        }^2},
\end{equation}
where $A$ and $C$ are constants to be determined by matching
conditions.

The solution for the geometric function $\lambda(r)$ is obtained
using (\ref{regularmet00})-(\ref{regularpress}) in
(\ref{edlrwss}), leading to
\begin{equation}\label{reglambda}
e^{-\lambda(r)}=1-\frac{2\tilde{m}(r)}{r},
\end{equation}
where the interior mass function $\tilde{m}$ is given by
\begin{eqnarray}\label{regularmass}
\tilde{m}(r)&=&m(r)-\frac{1}{\sigma}\frac{g(r)}{16\pi(1+Cr^2)(2+5Cr^2)^{11/10}}
\end{eqnarray}
with
\begin{eqnarray}
g(r)={-9\,
    C^2}\int_0^r \frac{r^2\, {\left( 2 + 5\, C\,
                  r^2 \right) }^{\frac{1}{10}}\, \left( -9 + 3\, C\,
            r^2 + 2\, C^2\, r^4 \right) }{{\left( 1 + C\, r^2 \right) }^2}\,
    dr
\end{eqnarray}
and $m(r)$ being the general relativity interior mass function,
given by the standard form
\begin{equation}
\label{regularmass2} m(r)=\int_0^r 4\pi
r^2{\rho}dr=\frac{3Cr^3}{4(1+Cr^2)},
\end{equation}
hence the total general relativity mass is obtained
\begin{equation}
\label{regtotmass} M\equiv m(r)\mid_{r=R}=\frac{3CR^3}{4(1+CR^2)},
\end{equation}
where $R$ is the radius of the distribution. A numerical analysis
on $g(r)$ shows that $g(r)>0$ for all $r\leq\,R$. Hence by
(\ref{regularmass}) we can see that the effective interior mass
function $\tilde{m}(r)$ is reduced by five dimensional effects.

Using (\ref{pp}) and (\ref{uu}) the interior Weyl functions are
written as
\begin{eqnarray}
\label{regP} {\cal P}(r)=&&\frac{1}{4 r^3\, {\left( 1 + C\, r^2
\right) }^4\, {\left( 2 + 5\, C\,
        r^2 \right) }^{\frac{21}{10}}}\left[2 {\left( 1 + C\, r^2 \right)
}^2\, \left(
          1 + 7\, C\, r^2 + 14\, C^2\, r^4 \right)g(r)\right.\nonumber\\ &&  \left.+ 3\, C^2\,
    r^3\, {\left( 2 + 5\, C\, r^2 \right) }^{\frac{1}{10}}\, \left( -18 -
          111\, C\, r^2 - 137\, C^2\, r^4 + 86\, C^3\, r^6 + 40\, C^4\,
        r^8 \right)\right]\nonumber\\,
\end{eqnarray}

\begin{eqnarray}\label{regU}
{\cal U}(r)=&&\frac{C}{{16
    r\, {\left( 1 + C\, r^2 \right) }^4\, {\left( 2 + 5\, C\,
            r^2 \right) }^{\frac{21}{10}}}}\left(
    8\, {\left( 1 + C\, r^2 \right) }^2\, \left(
          5 + 7\, C\, r^2 \right){g(r)}\right.\nonumber\\ &&  + \left.3\, C\,
    r\, {\left( 2 + 5\, C\, r^2 \right) }^{\frac{1}{10}}\, \left( -180 -
          660\, C\, r^2 - 509\, C^2\, r^4 + 62\, C^3\, r^6 + 55\, C^4\,
        r^8 \right)  \right)\nonumber\\\
\end{eqnarray}

It can be seen that $\lambda$, ${\cal U}$  and ${\cal P}$ do not
have exact expressions due to the presence of g(r). However this
feature does not impede the construction of a physically relevant
exact solution for the pressure and density, whose explicit brane
world expressions can be seen using matching conditions, where the
assumption of vanishing pressure will be dropped \cite{deru},
\cite{gergely2007}. As Schwarzschild is not the only possible
static exterior solution, we have many scenarios to consider. Let
us begin considering the simple Schwarzschild exterior solution
\begin{equation}
\label{SchwExt}
e^{\nu^{+}}=e^{-\lambda^{+}}=1-\frac{2\cal{M}}{r};\;\;\;\;\;\;\;{\cal
U}^{+}={\cal P}^{+}=0.
\end{equation}
The matching condition $[ds^2]_{\Sigma}=0$ at the stellar surface
$\Sigma$ yields
\begin{equation}
\label{RegmatchSchw1} A=(1-\frac{2\cal{M}}{R})(1+CR^{2})^{-3},
\end{equation}

\begin{eqnarray}\label{RegmatchSchw2}
\frac{2\cal{M}}{R}&=&\frac{2M}{R}-\frac{1}{\sigma}\frac{g(C)}{8{\pi}R(1+CR^2)(2+5CR^2)^{11/10}},
\end{eqnarray}
where
\begin{eqnarray}
g(C)\equiv g(r)|_{r=R}={-9\,
    C^2}\int_0^R \frac{r^2\, {\left( 2 + 5\, C\,
                  r^2 \right) }^{\frac{1}{10}}\, \left( -9 + 3\, C\,
            r^2 + 2\, C^2\, r^4 \right) }{{\left( 1 + C\, r^2 \right) }^2}\,
    dr.
\end{eqnarray}
Using (\ref{matchingfS}) it is found that $C$ must satisfy the
condition
\begin{eqnarray}
\label{RegmatchSchw355} 8{\pi}p(R)+\frac{1}{\sigma} \frac{g(C)\,
\left( 1 + 7\, C\, R^2 \right) }{8\, \pi \,
      R^3\, {\left( 1 + C\, R^2 \right) }^2\, {\left( 2 + 5\, C\,
              R^2 \right) }^{\frac{11}{10}}} =0,
\end{eqnarray}
which can be written as
\begin{eqnarray}
\label{RegmatchSchw3} \frac{9\,
      C\, \left( 1 - C\,
          R^2 \right) }{2 {\left( 1 + C\,
              R^2 \right) }^2} +\frac{1}{\sigma} \frac{g(C)\, \left( 1 + 7\, C\, R^2 \right) }{8\, \pi \,
      R^3\, {\left( 1 + C\, R^2 \right) }^2\, {\left( 2 + 5\, C\,
              R^2 \right) }^{\frac{11}{10}}} =0.
\end{eqnarray}
The equation (\ref{RegmatchSchw3}) shows clearly that $C$ cannot
be the general relativity value $C_0$, given by \footnote{A
numerical analysis shows that $g(C)\neq\,0$ for a wide range of
$R$.}
\begin{equation}
\label{RegC0} C_0=\frac{1}{R^2},
\end{equation}
which is found using the condition $p(R)=0$ in
(\ref{regularpress}). What the equation (\ref{RegmatchSchw3}) says
is that the general relativity value of $C$, namely $C_0$, has
been modified due to the buck effects by
\begin{equation}
\label{RegC} C=C_0+\delta(\sigma).
\end{equation}
In this sense $\delta(\sigma)$ represents the " bulk perturbation"
of the general relativity value of $C$. In order to find the bulk
contribution to $p$ and $\rho$ we need to find $C$, given by
(\ref{RegC}), satisfying (\ref{RegmatchSchw3}). Thus we have at
first order in $\sigma^{-1}$
\begin{equation}
\delta(\sigma)=\frac{1}{\sigma}\frac{2g(C_0)}{9\,\pi\,R^37^{11/10}}
\end{equation}
The pressure can thus be found expanding $p(C)$ around $C_0$
\begin{equation}
\label{ExpanP}
p(C_0+\delta)=p(C_0)+\delta{\frac{dp}{dC}}\mid_{C=C_0},
\end{equation}
which leads to
\begin{eqnarray} \label{RegPSchw} p(r)=\frac{9\,
    C_0\, \left( 1 - C_0\,
        r^2 \right) }{16\pi{\left( 1 + C_0\, r^2 \right)
        }^2}\,\,\,+&\underbrace{\frac{9}{16\pi}\frac{1-3C_0r^2}{(1+C_0r^2)^3}\delta(\sigma),}&
        \\ \nonumber
        &\delta\,p(\sigma)&
\end{eqnarray}
and by the same way the density is found to be
\begin{eqnarray}
\label{RegDSchw} \rho(r)=\frac{3\,
    C_0\, \left( 3 + C_0\,
        r^2 \right) }{16\pi {\left( 1 + C_0\, r^2 \right)
        }^2}\,\,\,+&\underbrace{\frac{3(3-C_0r^2)}{16\pi(1+C_0r^2)^3}\delta(\sigma)}&,
        \\ \nonumber
        &\delta\,\rho(\sigma)&\hspace{1cm}
\end{eqnarray}
where $\delta\,p(\sigma)$ and $\delta\,\rho(\sigma)$ represent the
bulk effects on $p(r)$ and $\rho(r)$ respectively. However at the
surface and for any arbitrary $R$ always we have
\begin{equation}
\label{RegPSchw2} p(R)=-\frac{9}{64\pi}\delta(\sigma) < 0.
\end{equation}
Hence the Schwarzschild exterior solution is incompatible with the
interior solution found here. Thus a different exterior solution
must be considered.

Using now the Reissner-N\"{o}rdstrom-like solution given in
\cite{dmpr}
\begin{equation}
\label{RegRNmet}
e^{\nu^+}=e^{-\lambda^+}=1-\frac{2\cal{M}}{r}+\frac{q}{r^2},
\end{equation}
\begin{equation}
\label{RegRNmet2} {\cal U}^+=-\frac{{\cal P}^+}{2}=\frac{4}{3}\pi
q\sigma\frac{1}{r^4},
\end{equation}
and considering the matching condition $[ds^2]_{\Sigma}=0$ at the
stellar surface $\Sigma$, we have
\begin{equation}
\label{RegmatchNR1}
A(1+CR^{2})^{3}=1-\frac{2\cal{M}}{R}+\frac{q}{R^2},
\end{equation}
\begin{eqnarray}\label{RegmatchNR2}
\frac{2\cal{M}}{R}&=&\frac{2M}{R}-\frac{1}{\sigma}\frac{g(C)}{8{\pi}R(1+CR^2)(2+5CR^2)^{11/10}}+\frac{q}{R^2},
\end{eqnarray}
and using (\ref{matchingf}) we obtain
\begin{eqnarray}
\label{qNR} q/R^2&=&-\frac{9\,
    C\, R^2\left( 1 - C\,
        R^2 \right) }{2{\left( 1 + C\, R^2 \right)
        }^2}-\frac{1}{\sigma}\frac{\left( 1 + 7\, C\,
              R^2 \right)g(C) }{
    8\pi\,R{\left( 1 + C\, R^2 \right) }^2\, {\left( 2 + 5\, C\,
            R^2 \right) }^{\frac{11}{10}}}
\end{eqnarray}

The constants $\cal{M}$ and $q$ are given in terms of $C$ through
equations (\ref{RegmatchNR2}) and (\ref{qNR}) respectively,
leaving $A$ and $C$ satisfying (\ref{RegmatchNR1}). As in the
previous case, $C$ and $A$ have a well definite general relativity
values, which are given respectively by the condition $p(R)=0$ in
(\ref{regularpress}) and evaluating (\ref{RegmatchNR1}) at
$\sigma^{-1}=0$
\begin{equation}
\label{RegmatchNR1GR} A_0(1+C_0R^{2})^{3}=1-\frac{2M}{R}.
\end{equation}
Comparing (\ref{RegmatchNR1}) from (\ref{RegmatchNR1GR}) it is
clear that the general relativity values of $A$ and $C$ have been
modified by bulk effects by
\begin{equation}
\label{RegAC}
C(\sigma)=C_0+\delta(\sigma);\,\,\,\,\,A(\sigma)=A_0+\varepsilon(\sigma).
\end{equation}

Using (\ref{RegAC}) in (\ref{RegmatchNR1}) we obtain
\begin{equation}
\label{RegmatchNR1Pert}
(A_0+\varepsilon)[1+(C_0+\delta)R^{2}]^{3}=1-\frac{2{\cal
M}}{R}+\frac{q}{R^2}.
\end{equation}
Evaluating the expressions (\ref{RegmatchNR2}) and (\ref{qNR}) at
$C=C_0 +\delta$ and keeping li\-neal terms in $\sigma^{-1}$, the
equation (\ref{RegmatchNR1Pert}) leads to
\begin{equation}
\label{dNR}
C(\sigma)=C_0+\frac{8}{3R^2}\left[\frac{1}{\sigma}\frac{g(C_0)}{16\pi\,R(7)^{11/10}}-8\varepsilon\right]+{\cal
O}(\sigma^{-1}).
\end{equation}
Hence $C(\sigma)$ is determinate if $\varepsilon$ is kept as a
free parameter which can be used in finding a physically
acceptable model. Thus it is possible to see bulk consequences on
$p$ and $\rho$ through (\ref{RegPSchw}) and (\ref{RegDSchw}). The
figure 1 shows the behaviour of the pressure in both the general
relativity and braneworld case. It can be seen that the five
dimensional gravity effects reduce the pressure deep inside the
distribution. However the situation changes for the exterior
layers, where the matching conditions lead to $p\neq 0$ at the
surface.

\section{Conclusions}

In the context of the astrophysics braneworld, a method to find
consistent solutions to Einstein's field equations in the interior
of a spherically symmetric, static and non uniform stellar
distribution with Weyl stresses was developed. During this process
a very important feature of the indefinite system
(\ref{ec1})-(\ref{con1}) was identified. When the simplest and
straightforward way to solve field equations is used, the general
relativity solution is lost. The source of this problem was
clearly identified, and a general solution for it was given
through the equation (\ref{edlrwss}), which gives the geometric
function $\lambda(r)$ written in a way that allows to see clearly
the bulk effects on the brane.

Keeping under control non-manifest bulk contributions when the
solution for $\lambda(r)$ is achieved on the brane, the loss of
the general relativity limit was avoided and then a new constraint
arose as a natural consequence of this method. A physical
interpretation of this constraint was given as the necessary
condition to regain general relativity. It was shown that this
constraint can be used as a natural tool which helps to close the
indefinite system (\ref{ec1})-(\ref{con1}) and therefore to
investigate bulk consequences on the brane.

An exact analytical solution on the brane was constructed by a
particular solution to the constraint (\ref{constraint}). This
particular solution represents a new constraint defined on the
brane, which was interpreted physically as the null consequence of
bulk corrections to $p$, $\rho$ and $\nu$ to produce anisotropic
effects on the brane.

By prescribing the temporal metric component $g_{00}$, an exact
interior solution to the pressure and density was found. This
represents a physically acceptable solution, namely, regular at
the origin, pressure and density definite positive, well definite
mass and radius, monotonic decrease of the density and pressure
with increasing radius, etc. It was shown that this solution is
incompatible with the Schwarzschild's exterior solution. Using the
Reissner-N\"{o}rdstrom-like metric given in \cite{dmpr}, the
effects of five dimensional gravity on pressure and density were
found through matching conditions. It was found that the bulk
gravity effect reduces the pressure deep inside the distribution,
but the situation changes for the exterior layers, as a direct
consequence of matching conditions.

As remarked earlier, it is well known that finding physically
relevant exact ste\-llar solutions in general relativity, even for
a perfect fluid, is not at all easy. Due to local and non-local
contributions from the bulk, the problem becomes even more
complicated. The exact solution for the physical variables $p(r)$
and $\rho(r)$ found in this paper was possible as a direct
consequence of the constraint (\ref{constraint2}). It is easy to
see through the field equations (\ref{ec1})-(\ref{ec3}) evaluated
at $\sigma^{-1}=0$ (general relativity) that the non-local
function $H(p,\rho,\nu)$ can be written as
\begin{equation}
\label{H2}
H(p,\rho,\nu)=\left(2G_2^2+G_1^1\right)|_{\sigma^{-1}=0}-8\pi 3p,
\end{equation}
which clearly correspond to an anisotropic term. In general
relativity the function $H(p,\rho,\nu)$ vanishes as a consequence
of the isotropy of the solution. However, in the braneworld case,
the condition $H(p,\rho,\nu)=0$ in general is not satisfied
anymore. There is not reason to believe that the modifications
undergone by $p$, $\rho$ and $\nu$, due to five dimensional
effects on the brane, do not modify the isotropic condition
$H(p,\rho,\nu)=0$. Therefore in general we have $H(p,\rho,\nu)\neq
0$ on the brane. However when the constraint $H(p,\rho,\nu)=0$ is
imposed to ensure the general relativity limit, all anisotropic
effects on the brane whose source are bulk corrections on
pressure, density and $\nu$, are eliminated, leaving only high
energy corrections for the geometric function $\lambda(r)$. Such
constraint not only ensures general relativity as a limit, but
also represents an enormous simplification for the function
$\lambda(r)$, which clearly helps when exact solutions are
investigated. Hence this constraint should be considered in
searching physically relevant exact solution on braneworld. Guided
by this, the following conjecture is made: {\it When a spherically
symmetric, static and non-uniform stellar distribution in the
context of the braneworld is considered, it is not possible to
ge\-nerate physically relevant exact analytical solutions on the
brane, unless bulk corrections on pressure, density and the
temporal metric component do not produce anisotropic
consequences.} This conjecture might help in the search of exact
and physically relevant solutions when a non-uniform distribution
is considered. Hence the role played by the inhomogeneousness as a
source of Weyl stresses in the interior could be studied. This is
currently been investigated.

\begin{figure}
  \includegraphics{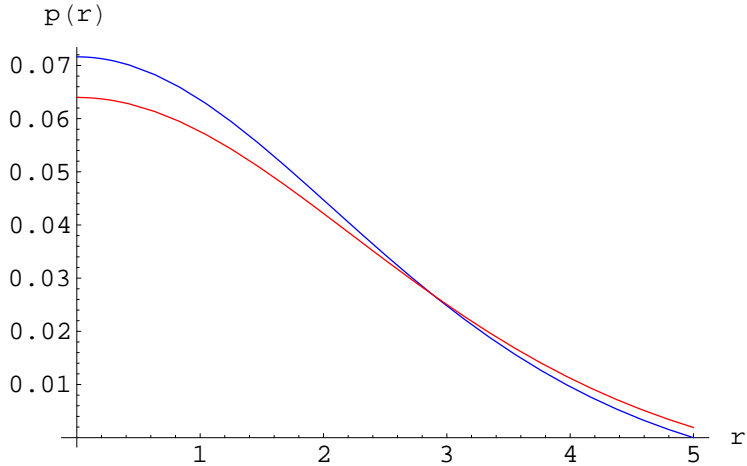}
\caption{Qualitative comparison of the pressure $p(r)$ in general
relativity
        ( $p(R)=0$ ) and the braneworld model ( $p(R)\neq 0$ ) with $R=5$.}
\label{fig:1}       
\end{figure}

\section*{Acknowledgments}

This work was supported by {\bf CDCH}. Grant:
CDCH-ICC03-0162-2007, and by {\bf FONACIT}. Grant: S2-2007001210.


\begin{thebibliography}{0}
\bibitem{rs} L. Randall and R. Sundrum,  \textit{Phys.Rev.Lett.} 83 (1999) 4690-4693.

\bibitem{chr} A. Chamblin, S.W. Hawking, H.S. Reall,\textit{ Phys.Rev. D}61 (2000) 065007.

\bibitem{sms} T. Shiromizu, K. Maeda and M. Sasaki, \textit{Phys.Rev. D}62 (2000) 024012 .

\bibitem{dmpr} N. Dadhich, R. Maartens, P. Papadopoulos, V. Rezania, \textit{Phys.Lett. B}487 (2000) 1-6

\bibitem{pert}J. Garriga and T. Tanaka, \textit{Phys.Rev.Lett.} 84 (2000) 2778-2781

\bibitem{gkr} S. Giddings, E. Katz and L. Randall, \textit{JHEP }0003 (2000) 023

\bibitem{dgp01} G. Dvali, G. Gabadadze, M. Porrati, \textit{Phys.Lett. B}485 (2000) 208-214

\bibitem{gm} C. Gordon and R. Maartens, \textit{Phys.Rev. D}63 (2001) 044022

\bibitem{dk} N. Deruelle and J. Katz, \textit{Phys.Rev. D}64 (2001) 083515

\bibitem{germ2} R. Maartens, \textit{Reference Frames and Gravitomagnetism}, ed. J Pascual-Sanchez et al. (World Sci., 2001), p93-119 [gr-qc/0101059.]

\bibitem{kmp03} G. Kofinas, R. Maartens and E. Papantonopoulos, \textit{JHEP} 0310 (2003) 066

\bibitem{anaaeg04} A.N. Aliev and A.E. Gumrukcuoglu, \textit{Class.Quant.Grav}. 21 (2004) 5081-5096

\bibitem{asmnm} A. S. Majumdar, N. Mukherjee, \textit{Int.J.Mod.Phys.D} 14 (2005) 1095

\bibitem{cosmox} R. A. Brown, R. Maartens, E. Papantonopoulos, V.Zamarias, \textit{JCAP} 0511 (2005) 008

\bibitem{maartRev2004} R. Maartens, \textit{Living Rev.Rel.} 7 (2004) 7

\bibitem{bdel00} P. Binetruy, C. Deffayet,U. Ellwanger, D. Langlois, \textit{Phys.Lett. B} 477 (2000) 285-291

\bibitem{kmdw00} K. Maeda, D. Wands, \textit{Phys.Rev.D }62 (2000) 124009

\bibitem{maart00} R. Maartens, \textit{ Phys.Rev.D} 62 (2000) 084023

\bibitem{lang01} D. Langlois, \textit{Phys.Rev.Lett.} 86 (2001) 2212-2215

\bibitem{accfp01} A. Campos, C. F. Sopuerta, \textit{Phys.Rev.D} 63 (2001) 104012

\bibitem{chm01} Chiang-Mei Chen, T. Harko, M. K. Mak, \textit{Phys.Rev.D} 64 (2001)

\bibitem{coley02} A. Coley, \textit{Phys.Rev.D} 66 (2002) 023512 044013

\bibitem{bmw02} H. A. Bridgman, K. A. Malik, D. Wands, \textit{Phys.Rev.D} 65 (2002) 043502

\bibitem{ekiri05}  E. Kiritsis, \textit{JCAP} 0510 (2005) 014

\bibitem{psant06} P. S. Apostolopoulos, N. Tetradis, \textit{Phys.Lett.B} 633 (2006) 409-414

\bibitem{cbkt07} C. Bogdanos, K. Tamvakis ,\textit{Phys.Lett.B} 646 (2007) 39-46

\bibitem{cosmo04} P. Brax, C. van de Bruck and A. C. Davis, \textit{Rept.Prog.Phys.} 67 (2004) 2183-2232

\bibitem{germ} C. Germani, R. Maartens, \textit{Phys.Rev.D}64 (2001) 124010

\bibitem{deru} N. Deruelle, {\it Stars on branes: the view from the brane} gr-qc/0111065.

\bibitem{twise}  T. Wiseman, \textit{Phys.Rev.D} 65 (2002) 124007

\bibitem{kpp02} G. Kofinas, E. Papantonopoulos and I. Pappa, \textit{Phys.Rev.D} 66 (2002) 104014

\bibitem{ppz02} G. Kofinas, E. Papantonopoulos and V. Zamarias, \textit{Phys. Rev.D}66 (2002) 104028

\bibitem{vw} M. Visser, D.L. Wiltshire, \textit{Phys.Rev.D}67 (2003) 104004

\bibitem{anaaeg2005} A.N. Aliev and A.E. Gumrukcuoglu, \textit{Phys.Rev.D} 71 (2005) 104027

\bibitem{CGKM} S. Creek, R. Gregory, P. Kanti, B. Mistry, \textit{Class.Quant.Grav}. 23 (2006) 6633-6658

\bibitem{baff} B. Ahmedov, F. Fattoyev {\it Magnetic Fields of Spherical Compact Stars in Braneworld}, gr-qc/0608039.

\bibitem{twism} T. Wiseman, \textit{Class.Quant.Grav.} 19 (2002) 3083-3106

\bibitem{bgm} M. Bruni, C. Germani and R. Maartens \textit{Phys.Rev.Lett.} {\bf 87} (2001) 231302

\bibitem{ndsgg} N. Dadhich, S. G. Ghosh, \textit{Phys.Lett.B} 518 (2001) 1-7

\bibitem{mgnd} M. Govender, N. Dadhich, \textit{Phys.Lett.B}538 (2002) 233-238

\bibitem{kp2004} G. Kofinas and E. Papantonopoulos, \textit{JCAP} {\bf 0412} (2004) 011.

\bibitem{spal} S. Pal, \textit{Phys.Rev.D}74 (2006) 124019

\bibitem{gergely2007} L\'aszl\'o \'A. Gergely, \textit{JCAP}02(2007)027

\bibitem{kramer} D. Kramer, H. Stephani, E. Herlt and M. MacCallum, {\it Exact Solutions of Einstein's Field Equations}(Cambridge University Press, Cambridge, 1980).

\bibitem{DK98} M.S.R. Delgaty and Kayll Lake \textit{Comput.Phys.Commun}.115 395-415(1998).

\bibitem{flobo07} Francisco S.N. Lobo, \textit{Phys.Rev.D}75 (2007) 064027.

\bibitem{shtanov07} A. Viznyuk and Y. Shtanov, \textit{Phys.Rev. D}76 (2007) 064009

\bibitem{gergely2005} L\'aszl\'o \'A. Gergely, \textit{Phys. Rev. D} 71 (2005) 084017, Erratum: ibid 72 (2005) 069902

\bibitem{gergely2006} L\'aszl\'o \'A. Gergely, \textit{Phys. Rev. D} 74 (2006) 024002

\bibitem{gergely2007-2} L\'aszl\'o \'A. Gergely, I. K\'ep\'{\i}r\'o, \textit{JCAP} 0707 (2007) 007

\end{thebibliography}
\end{document}